\documentclass[aps,twocolumn,showpacs,superscriptaddress, amsmath,amssymb]{revtex4}

\usepackage{graphicx}% Include figure files
\usepackage{dcolumn}% Align table columns on decimal point
\usepackage{bm}% bold math
\usepackage{graphics}% Include figure files
\usepackage{subfigure}
\usepackage{amsmath}
\usepackage[T1]{fontenc}
\usepackage{color}
\usepackage{bm}

\begin{document}

\title{Compaction of mixtures of rigid and highly deformable particles: \\
a micro-mechanical model}
\author{Manuel C\'ardenas-Barrantes}
\email{manuel-antonio.cardenas-barantes@umontpellier.fr}
\affiliation{LMGC, Universit\'e de Montpellier, CNRS, Montpellier, France}
\author{David Cantor}
\email{david.cantor@polymtl.ca}
\affiliation{Department of Civil, Geological and Mining Engineering, Polytechnique Montr\'eal, Qu\'ebec, Canada}
\author{Jonathan Bar\'es}
\email{jonathan.bares@umontpellier.fr}
\affiliation{LMGC, Universit\'e de Montpellier, CNRS, Montpellier, France}
\author{Mathieu Renouf}
\email{mathieu.renouf@umontpellier.fr}
\affiliation{LMGC, Universit\'e de Montpellier, CNRS, Montpellier, France}
\author{Emilien Az\'ema}
\email{emilien.azema@umontpellier.fr}
\affiliation{LMGC, Universit\'e de Montpellier, CNRS, Montpellier, France}
\date{\today}

\begin{abstract}
We analyze the isotropic compaction of mixtures composed of rigid and deformable incompressible particles by the non-smooth contact dynamics approach (NSCD).
The deformable bodies are simulated using a hyper-elastic neo-Hookean constitutive law by means of classical finite elements.
For mixtures that varied from totally rigid to totally deformable particles, we characterize the evolution of the
packing fraction, the elastic modulus, and the connectivity as a function of the applied stresses when varying inter-particle coefficient of friction.
We show first that the packing fraction increases and tends asymptotically to a maximum value $\phi_{max}$,
which depends on both the mixture ratio and the inter-particle friction.
The bulk modulus is also shown to increase with the packing fraction and
to diverges as it approaches $\phi_{max}$.
From the micro-mechanical expression of the granular stress tensor,
we develop a model to describe the compaction behavior as a function of the applied pressure,
the Young modulus of the deformable particles, and the mixture ratio.
A bulk equation is also derived from the compaction equation.
This model lays on the characterization of a single deformable particle under compression together with a
power-law relation between connectivity and packing fraction.
This compaction model, set by well-defined physical quantities, results in outstanding predictions
from the jamming point up to very high densities and allows us to give a direct prediction of $\phi_{max}$ as a function of both the mixture ratio and the friction coefficient.
\end{abstract}

\maketitle

%------------------------------------------------------------------------------
\section{Introduction}

Mixtures of particles with different bulk properties are the constitutive element of many materials playing a crucial role in many natural and industrial processes.
Among these materials are biological tissues composed of soft cells \cite{Mauer2018,Wyatt2015_Emergence,bi2016_prx}, foams \cite{Bolton1990_Rigidity,Katgert2010_Jamming}, suspensions \cite{Dijksman2017,Stannarius2019,bares2020_Transparent,brujic2003_3D}, clayey materials \cite{rua2019_Modeling,iliecs2017_Soil}, and any sintered material \cite{Cooper1962,Heckel1961,Kawakita1971,Kim1987} as ceramic, metal or pharmaceutical pills to name a few.
In an engineering context, recent emerging issues have led to the design of new materials in the form of a mixture of soil particles
with rubber pieces (made from discarded tires). Such composite material exhibits new and
fascinating mechanical properties such as better stress relaxation \cite{Indrarantna2019_Use,ASTM2008_Standard,Tsoi2011_Mechanical,Khatami2019_The}, seismic isolation \cite{Tsang2008_Seismic,Senetakis2012_Dynamic,Mavronicola2010_Numerical,Tsiavos2019_A}
and foundation damping \cite{Indrarantna2019_Use,Anastasiadis2012_Small,Mashiri2015_Shear,Khatami2019_The} while reducing the weight of the structures or increasing the packing fraction of the granular composites.
The range of applications for rigid-deformable composites is potentially broad and opens the door to an extensive field of fundamental topics that are still poorly studied.

The mechanical behavior of a packing of deformable particles mainly depends on the ability of the particles to both, rearrange (sliding or rolling) and change of their shape (related to the Young's modulus and the Poisson's ratio of the particles).
For example, at the outset of compression, the granular assembly tends to the jammed state mainly by inner particle rearrangements until a mechanical equilibrium is reached withstanding the imposed loading.
Once jammed, if the compression continues, the particle deformation is the main mechanism that permits the system to find a new mechanical equilibrium.
Hence, the complexity of rigid-deformable mixtures arises from geometrical and mechanical dissimilarities between particles, leading to possible rearrangements even after the jamming point.

The compaction mechanism of soft granular matter, especially beyond the jamming point, is a broad issue increasingly studied in the literature
both experimentally \cite{Liu2018_Compression,Vu2019,Platzer2018,Lee2007_Behavior,Youwai2003_Strength,Asadi2018_Discrete} and numerically through discrete element methods \cite{Asadi2018_Discrete,Gong2019_Direct,Jonsson2019_Evaluation,Agnolin2008_On,Lopera2017_Micromechanical,Boromand2018}, meshless approaches \cite{Platzer2018,Mollon2018,Nezamabadi2019} or coupled
finite-discrete element methods \cite{Procopio2005,Huang2017,Abdelmoula2017,Wang2020_Particulate}. Still, even if many descriptions of these systems have been made, an understanding of the main mechanisms and theoretical framework is missing. Indeed,
a large number of equations trying to link the confining pressure $P$ to the packing fraction $\phi$ (i.e., the ratio between the volume of the particles $V_p$ over the volume of the box $V$) have been proposed, but most of them based on empirical strategies.

One of the first constitutive equation was proposed by Walker in 1923 \cite{Walker1923_The}. This states that the packing fraction, $\phi$, is proportional to  the logarithm of the pressure ($\ln P$). This model involves two fitting constants which have been later correlated to
an equivalent Young's modulus or yield strength \cite{Balshin1938_Theory}.
Shapiro and Kolthoff \cite{Shapiro1947_Studies}, followed by Konopicky \cite{Konopicky1948_Parallelitat} and Heckel \cite{Heckel1961},
had a different approach and assumed the proportionality between the porosity ($1-\phi$) and the packing
fraction increment over the stress increment ($d\phi/dP$).
They proposed that $P \propto \ln(1-\phi)$ with, again, two fitting constants related to the powder properties. Later, Carroll and Kim  \cite{Carroll1984} justified this equation by correlating the loss of void space in the packing and the collapse of a spherical cavity within an elastic medium.
Many other compaction equations have been proposed in the literature \cite{Kawakita1971,Shapiro1993_Comp, Panelli2001, Denny2002,Montes2010}. However, like the previous ones, they relate linearly the logarithm of the packing fraction and a polynomial function of $P$,
with two or three fitting constants.
Some models also introduce a maximum packing fraction $\phi_{max}$, which  depends on the
grains' properties (shape, size, friction coefficient,...) in the form of $P \propto \ln(\phi_{max}-\phi)$ \cite{Secondi2002_Modelling}.
Recently, double logarithmic functions have also been  proposed by
Ge et al. \cite{Ge1995_A}, Zhang et al. \cite{Zhang2014} and  W{\"u}nsch et al. \cite{Wunsch2019_A}. They link linearly $\ln P$ to $\log \ln \phi$.
Less usual, non-linear equations have also been proposed by some authors \cite{Cooper1962,VanderZwan1982,Gerdemann2011_Compaction,Nezamabadi2019},
linking a functional form of $P$ to a functional form of $\phi$.
Those models still introduce fitting constants and most of them do not consider a maximum packing fraction.

An extensive list of equations mostly designed for metal or pharmaceutical
powders compactions is reviewed in \cite{Parilak2017,Comoglu2007,Machaka2015_Analysis}.
Although the previous models provide acceptable predictions on specific cases, their limitations are due to ($i$) the use of parameters with not a clear physical meaning, ($ii$) the lack of physical derivation, and, for many among them, ($iii$) the limitation to a single solid granular phase.

For assemblies of two distinct solid granular phases (i.e., for binary mixtures), the, so far, adopted strategies consist in using existing compaction equations for a single granular phase and free-parameter fitting \cite{Popescu2018_Compaction,Wang2020_Particulate}.
However, and to our best knowledge, the first attempt to predict the compaction behavior of mixtures of rigid-deformable particles
can be attributed to Platzer et al. \cite{Platzer2018}, who studied mixtures of sand with rubber particles.
They introduced an equation involving four parameters and deduced from the assumption that the empty space is filled as a first-order differential equation of the applied pressure.
Nevertheless, the authors mentioned that their model provides fair predictions for low pressures and a ratio of rigid to deformable
particles below 50 \%, but it loses its accuracy for high pressure.
It is worth noting that, as discussed later in Appendix \ref{A_Platzer}, their model is a more general form of previous models.

The large number of constitutive equations aiming at describing the evolution of compressed granular materials shows that there is currently no consensus on the micro-mechanisms taking place during
the compaction.
A more proper description and modeling of the compaction process should consider the multi-contact
nature of the assembly together with the deformability of the particles.

In this paper, we analyze the compaction behavior of mixtures of rigid and deformable particles by using a coupled discrete element and finite element method: the non-smooth contact dynamics (NSCD) approach.
We study the effect of the proportion of rigid-deformable particles in the mixture and the interparticle friction
on the compaction evolution and elastic properties beyond
the jamming point. Starting from the micro-mechanical definition of the granular stress tensor, we introduce an analytical
model for the compaction behavior accounting for the evolution of particle connectivity,
the applied pressure, the packing fraction, and the mixture ratio (i.e., the proportion of rigid-deformable particles in the assembly).
Our model accurately predicts the sample density ranging from the granular jamming point up to
high packing fractions for all mixture ratios and friction coefficients.
This model extends to binary mixtures our previous work developed for an assembly of only deformable particles \cite{Cantor2020_Compaction}.
As a natural consequence, the bulk modulus evolution and the maximum density a mixture can reach are also deduced.

The paper is structured as follows: in \textbf{Section \ref{Sec_Num}},
we briefly describe the numerical method used in the simulations, the construction of the samples, and
the procedure followed during the compaction.
In \textbf{Section \ref{Sec_Macro}}, the evolution of the packing fraction and the bulk properties beyond the jamming are analyzed
as a function of the applied pressure for different values of the mixture ratio and friction.
In this section, a discussion is also proposed regarding
the approximation given by some existing models to our results.
In \textbf{Section \ref{Sec_micromodels}}, a micro-mechanical model of compaction is
presented and validated. Finally, conclusions and perspectives are discussed in \textbf{Section \ref{Sec_conclu}}.

\section{Numerical procedures}
\label{Sec_Num}

\subsection{A coupled discrete-finite element method}
The simulations are carried out by means of the Non-Smooth Contact Dynamics (NSCD) method originally developed by Moreau and Jean \cite{Moreau1994_Some,Jean1999}.
The NSCD is the extension of the Contact Dynamic (CD)
method \cite{Jean1999,Dubois2018} to deformable bodies.
The CD method is a discrete approach for the simulation of granular dynamics considering contact laws with non-penetrability and dry friction between particles. In particular, the CD method does not require elastic repulsive potential or smoothing of the friction law for the determination of forces.
It is hence unconditionally stable and well suited to the simulation of large packings
composed of frictional particles of any shape.
An iterative and parallelized algorithm of resolution is used \cite{Renouf2004},
to simultaneously find the contact forces and changes of the momentum of each grain over time steps.

The deformable particles are discretized via classical finite element techniques, the degrees of freedom are the coordinates of the nodes.
We used an implementation of the NSCD on the free, open-source simulation
platform LMGC90 \cite{LMGC90_web}, developed in Montpellier and capable of modeling a collection of deformable or non-deformable particles. More details about the mathematical formulation and the implementation  of this numerical method are given in \cite{Jean1999,Dubois2018}.

\subsection{Packing construction, isotropic compression, and dimensionless parameters}

\begin{figure}
    \centering
    \includegraphics[width=0.7 \linewidth]{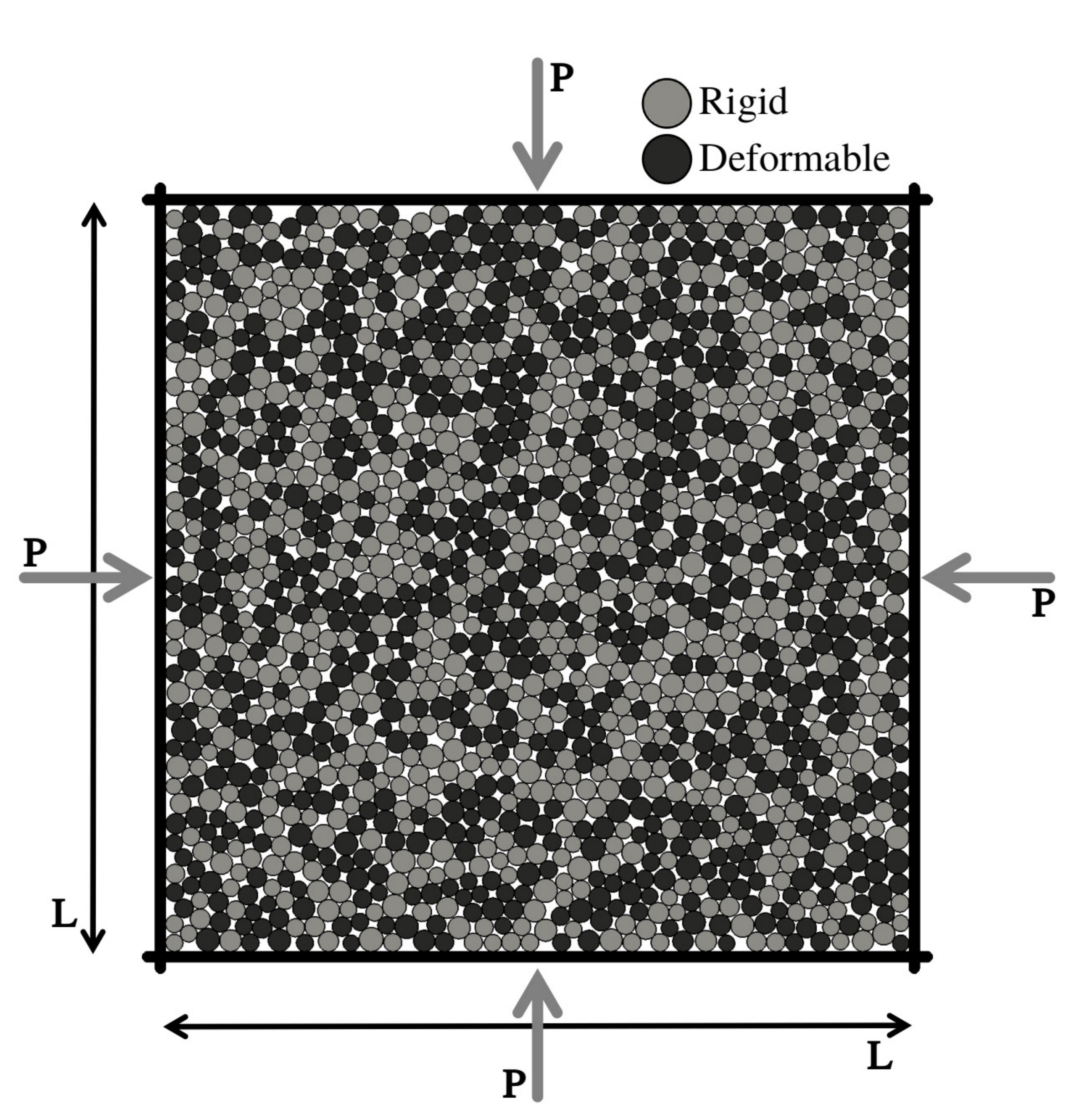}
    \caption{(Color online)  Scheme of the compression test for $\kappa = 0.50$. A collection of deformable and rigid particles
are prescribed inside an initially squared 2D frame and compressed in a quasistatic manner with an imposed pressure. $P$ is the applied pressure and $L$ is the size of the square box. }
    \label{fig:Scheme_compre}
\end{figure}

All samples are prepared according to the same protocol.
First, $N_p = 1500$ rigid disks are randomly placed into a square box of initial length $L_0$ by simple geometrical rules in
order to build a dense system \cite{Visscher1972}.
A weak size polydispersity is considered by varying the diameter $d$ of the disks in the range of $[0.8\langle d\rangle ,1.2\langle d\rangle]$
with a uniform distribution of the particle volume (area in 2D) fractions and $\langle d\rangle$ the mean diameter.

Second, a volume $\kappa V_p$ of rigid disks is homogeneously replaced by deformable disks meshed with 92
triangular similar-size elements, with $\kappa$ the mixture ratio
varying from $0.2$ (packing composed of 20\% of deformable particles) to $1$ (packing of only deformable particles).
All deformable particles are assumed to have the same isotropic
neo-Hookean incompressible constitutive law \cite{Rivlin1948}.
We use a constant Poisson's ratio equals to $0.495$ and a Young modulus $E$. Plane-strain conditions are also assumed.

Finally, the packings are isotropically compressed by gradually, and quasistatically,
applying a stress $P$ on the boundaries, as shown in Fig. \ref{fig:Scheme_compre}.
A set of loading steps are undertaken targeting stable values of applied pressure and packing fraction.
For a given pressure $P$, a stable state is reached once the variations of the packing fraction remained below 0.01\%.
The friction with the walls and the gravity are set to $0$ to avoid force gradients in the sample.

The relevant dimensionless control parameters for disks under pressure $P$ are the reduced pressure $P/E$ \cite{Agnolin2007b,Agnolin2007c} and the inertia parameter $I$ \cite{Andreotti2013} to assess how dynamic the tests are.
$I$ is defined as $\dot{\gamma} \langle d \rangle \sqrt{\rho/P}$,
where $\dot{\gamma} = v/L_0$ with $v$ the velocity of the walls, and $\rho$ the particle density. In all our simulations, $I$ remained below $10^{-4}$ so the particle-to-particle interaction and the particles' bulk rapidly damped the kinetic energy and elastic waves had little influence on the particle reorganization.
Note that as $P/E\rightarrow 0$, we have $\phi \rightarrow \phi_0$ with $\phi_0$ the packing fraction at the corresponding jammed state for the rigid assembly of particles.

\begin{figure}
\centering
\includegraphics[width=\linewidth]{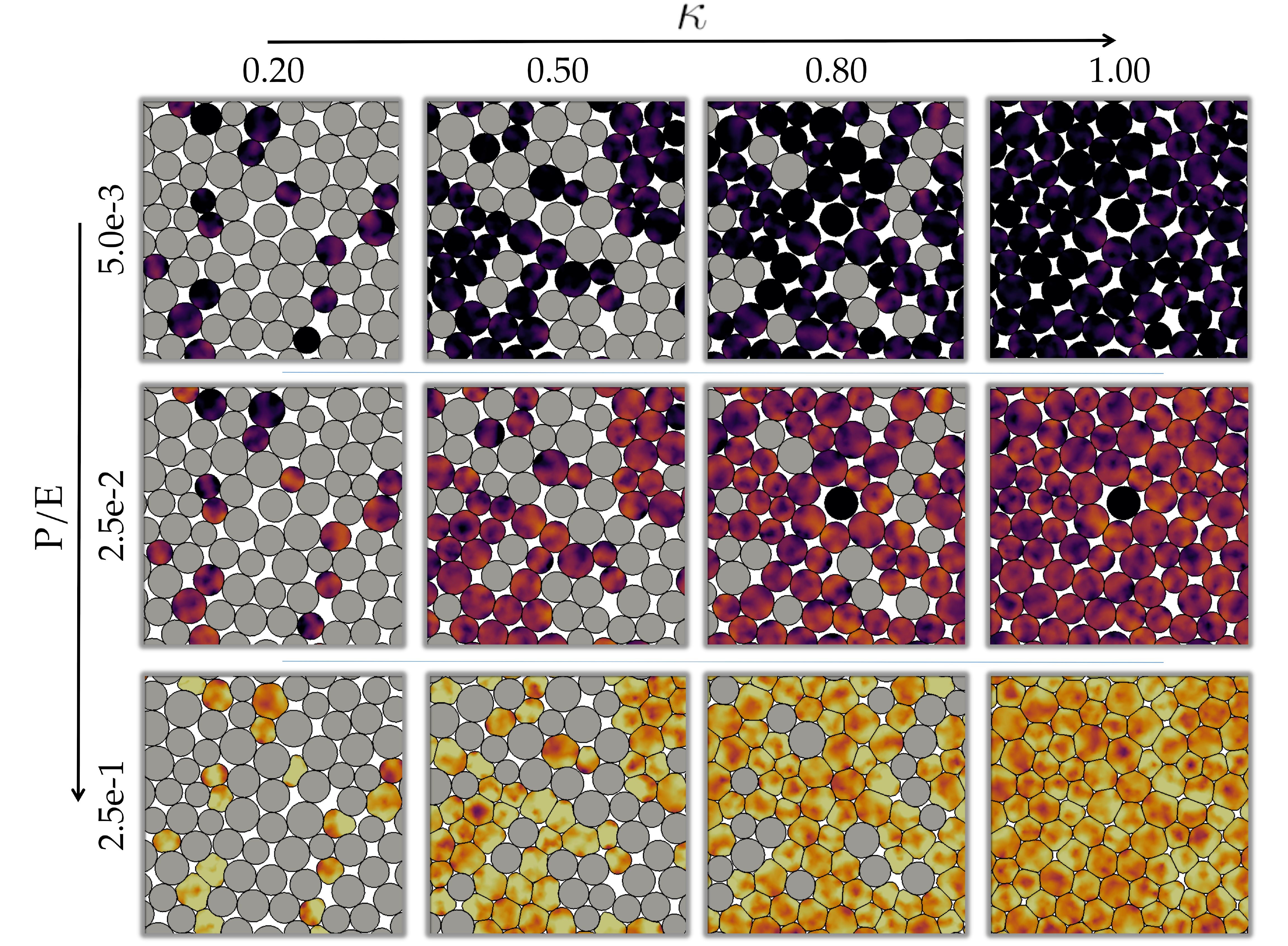}
\caption{(Color online) Close-up views on some of the samples for different mixture ratio $\kappa$ and the reduced pressure $P/E$.
Here, friction is fixed to $0.2$.
The rigid particles are shown in grey and the color intensity for the deformable ones
is proportional to the volumetric deformation within the particles. }
\label{fig:Snap-shots}
\end{figure}

We performed a large number of isotropic compression tests for a broad set of combinations of the
mixture ratio, the coefficient of friction, and the reduced pressure $P/E$.
The mixture ratio $\kappa$ was varied in the set $[0.2, 0.5, 0.8,1.0]$ for two distinct values of coefficient of friction $\mu_s=0.0$ and $\mu_s=0.2$.
For $\kappa=1$, we also include simulations for increasing coefficient of friction ($\mu_s=0$ to $0.8$).
The reduced pressure  $P/E\sim$ was varied from  $10^{-5}$ to $P/E\sim 5\cdot10^{-1}$.
The packings are shown in Fig. \ref{fig:Snap-shots} for different values of $\kappa$ and increasing
stable values of $P/E$ at $\mu_s=0.2$.

\section{Packing fraction and bulk properties}
\label{Sec_Macro}

\subsection{Numerical results}
\begin{figure}
    \centering
   \includegraphics[width=\linewidth]{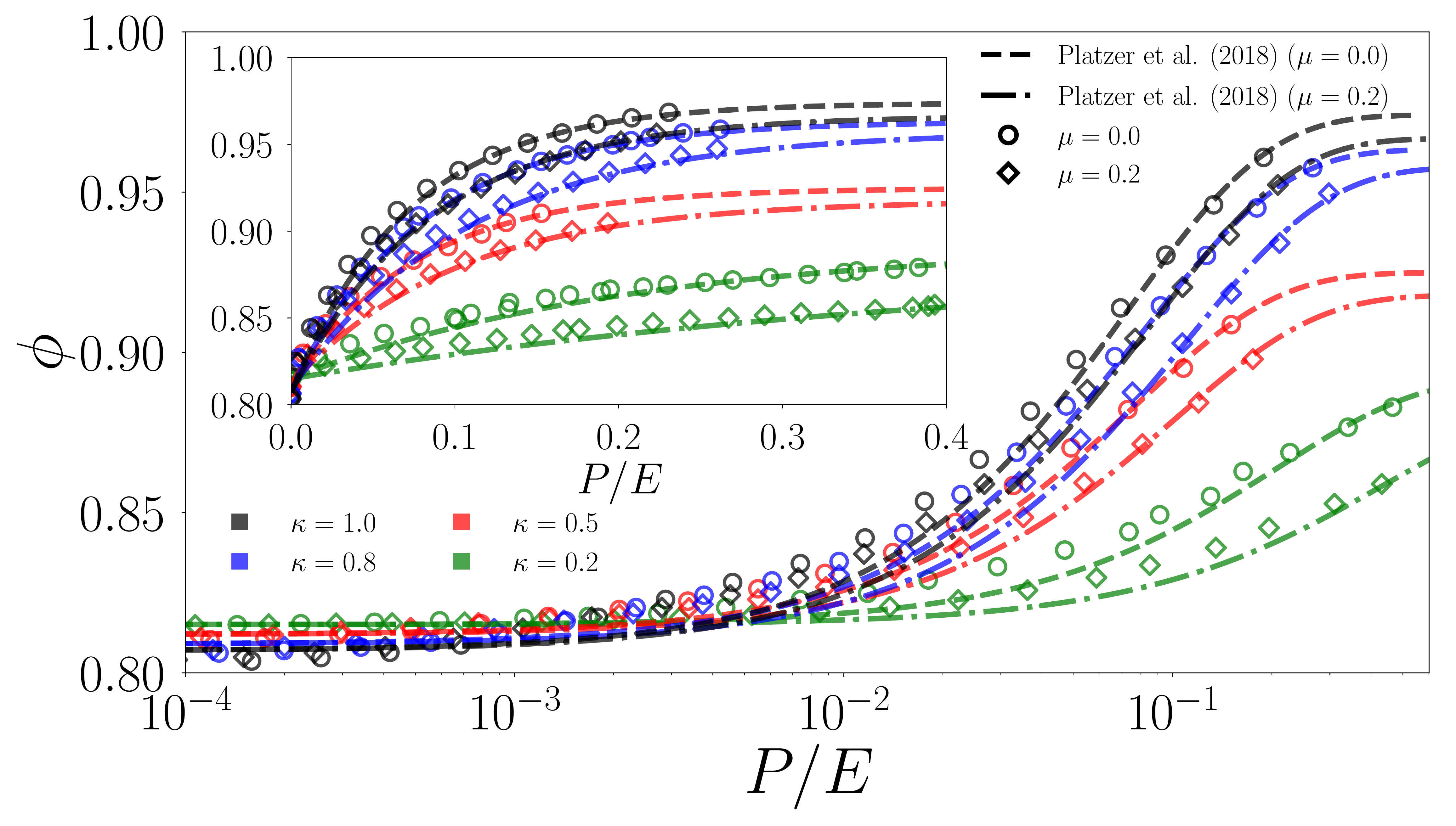}(a)
   \includegraphics[width=\linewidth]{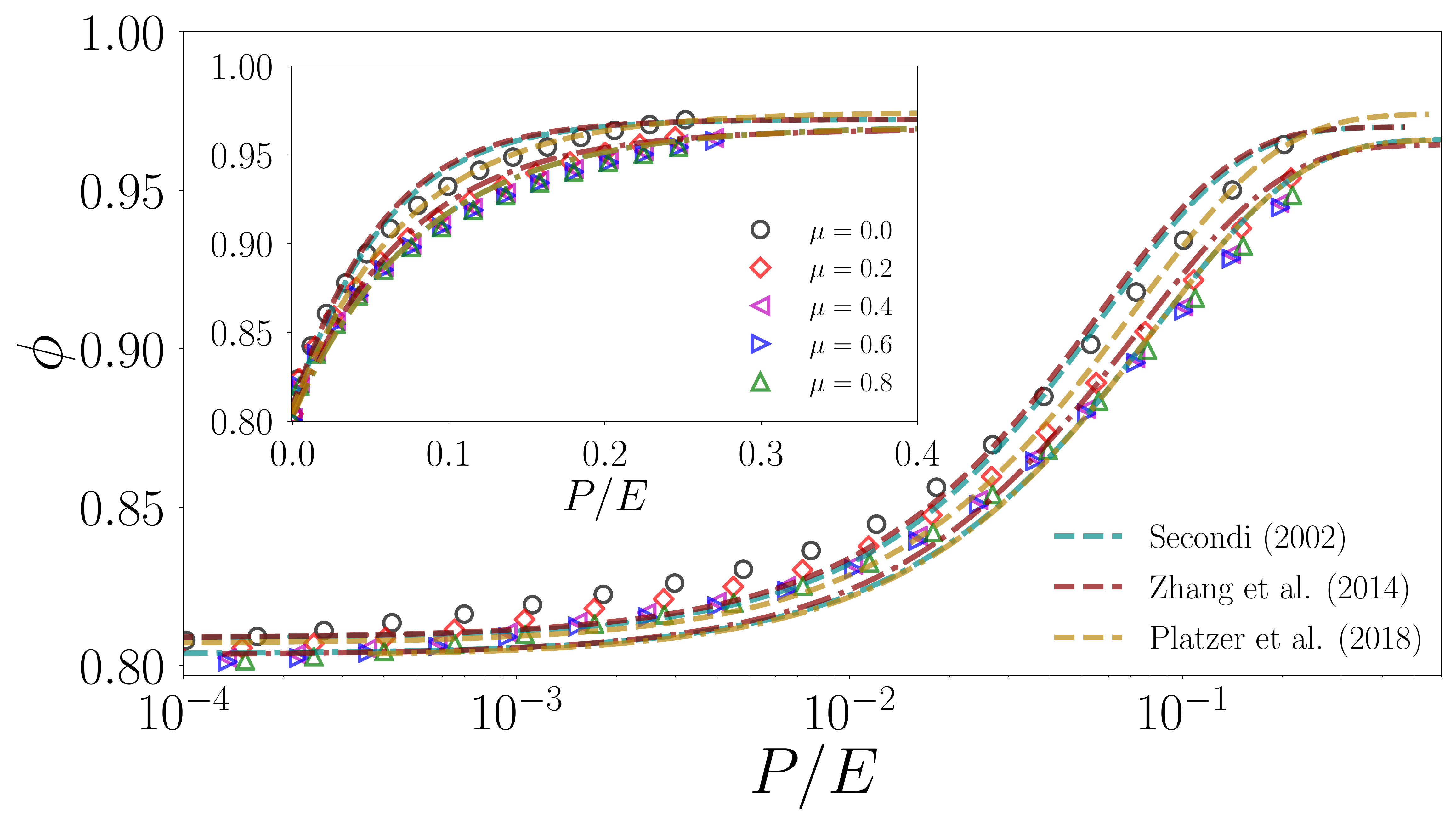}(b)
   \caption{(Color online)  Packing fraction $\phi$ as a function of $P/E$ for (a) rigid-deformable particles assemblies with $\kappa \in[0.2,...,1]$ and $\mu_s\in\{0,0.2\}$, and for  (b) completely deformable particle assemblies  (i.e., $\kappa=1$) with $\mu_s\in\{0, 0.8\}$. Main panels are in lin-log scale while insets show curves in lin-lin scale. Numerical data (symbols) are shown with fits from simplified equation of Platzer et al. (Eq. (\ref{Eq:Platzer})) for rigid-deformable particles and the equations of Secundi (Eq. (\ref{Eq:Secondi})) and Zhang (Eq. (\ref{Eq:Zhang})) for the fully deformable systems (dashed lines). }
    \label{fig:phi_p}
\end{figure}

Figure \ref{fig:phi_p} shows the evolution of $\phi$ as a function of $P/E$ for
rigid-deformable particle assemblies with $\kappa \in[0.2,0.5,0.8,1]$
and $\mu_s\in\{0,0.2\}$ (a), together with simulations fixing $\kappa=1$ and varying $\mu_s\in[0,0.2,0.4,0.6,0.8]$ (b).

For all cases, the evolution curves have the same general trend regardless of $\kappa$ and $\mu_s$. More particularly,
the packing fraction first increases with $P/E$ from $\phi_0$  and then tends asymptotically to a maximum packing
fraction $\phi_{max}$. We note that both $\phi_0$ and $\phi_{max}$ slightly decline as the local friction increases.
It is explained by the reduction of the particle rearrangements due to friction, as discussed in previous
studies \cite{Hecke2009_Jamming,Vu2020_compaction,Majmudar2007_Jamming}.

In assemblies of rigid-deformable particles, $\phi_{max}$ decreases as $\kappa$ tends towards $0$.
On the same curves, we plot some of the compaction models found in the literature fitting our numerical results for $\mu_s\in\{0,0.2\}$ for all mixtures and $\mu_s\in\{0,0.8\}$ at $\kappa=1$.
Further discussion upon these models is presented later in Sec. \ref{Discussion_models}.

\begin{figure}
    \centering
    \includegraphics[width=\linewidth]{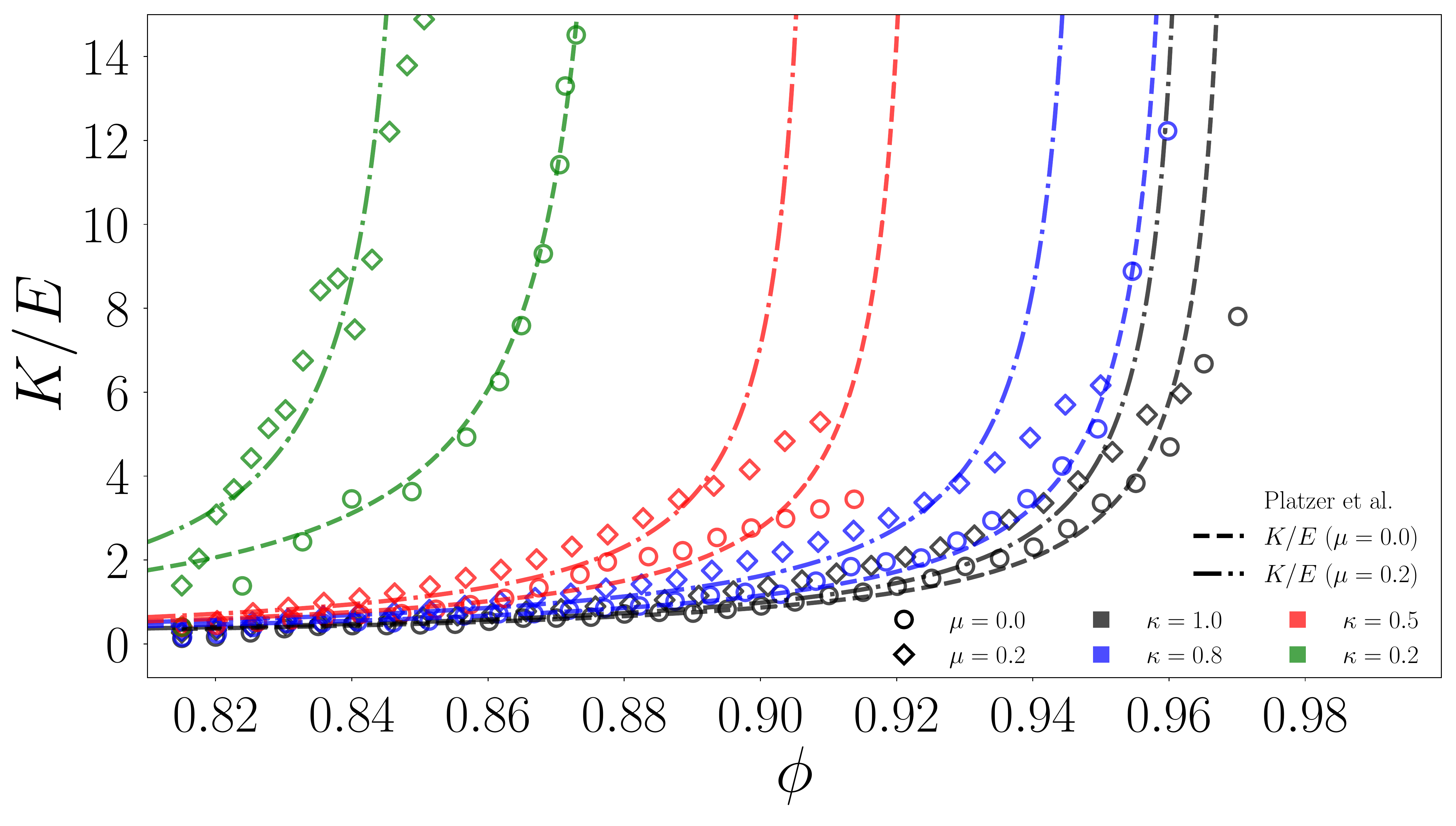}
    \caption{(Color online) Evolution of the bulk modulus $K$ normalized by the Young modulus $E$ as a
    function of the packing fraction $\phi$ for simulations with all values of $\kappa$ and for $\mu_s=0$ and $0.2$.
    The dashed lines show the bulk modulus computed from the Platzer et al. equation (Eq. (\ref{Eq:Platzer})).}
    \label{fig:BULK}
\end{figure}

It is also interesting to analyze the elastic properties of the assemblies depending on the values of
the mixture ratio and friction coefficient.
We define the bulk modulus as:
\begin{equation}
\label{Eq_Modulus}
K(\phi)= \frac{dP}{d\phi} \cdot \frac{d\phi}{d\varepsilon_v},
\end{equation}
with $\varepsilon_v = -\ln(\phi_0/\phi)$ the
macroscopic cumulative volumetric strain. Figure \ref{fig:BULK} shows the evolution of $K(\phi)$ as a function of $\phi$
for all values of $\kappa$ and $\mu_s$, measured in our simulations and computed using the derivative of compaction
equations used to fit the compaction curves (see the discussion in Sec. \ref{Discussion_models}).
We observe that $K$ increases with $\phi$ and diverges as the packing fraction tends to $\phi_{max}$. This comes from the fact that the assembly of grains starts to behave as a non-deformable solid.
We also note that, regardless of $\kappa$, the coefficient of friction has little influence on the macroscopic bulk modulus for the small deformation domain. However, its effect slightly increases for large strain values.

\subsection{Discussion}
\label{Discussion_models}
In this section, we discuss the approximation of our numerical data using some of the equations found in the literature.
For assemblies of only deformable particles (i.e., for $\kappa=1$), we display the model of Secondi  \cite{Secondi2002_Modelling} and the model of Zhang et al. \cite{Zhang2014}.
Secondi proposes an equation in the form of:
\begin{equation}
P^n  = -A_1\ln\Big[ \frac{\phi_{max}-\phi}{\phi_{max}-\phi_0}\Big],
\label{Eq:Secondi}
\end{equation}
where $n$ and $A_1$ are coupled parameters assumed to control the hardening and plasticity of the assembly at the macroscopic scale.
The use of this equation is motivated by the fact that it seems to generalize many other previously stated equations. For instance, assuming that $n=\phi_{max}=1$, we get the Heckel's equation \cite{Heckel1961}.
For $n=1$, we find the one proposed by Heuberger \cite{Heuberger1950} and Ballhausen \cite{Ballhausen1951}, while for $\phi_{max}=1$, we obtain the equation of Parilak et al. \cite{Panelli2001}.
Similarly, Panely's equation appears for $n=0.5$ and $\phi_{max}=1$ \cite{Parilak2017}.
In contrast, the equation proposed by Zhang et al. states that:
\begin{equation}
\label{Eq:Zhang}
\log P = m\log\ln \left[ \frac{(\phi_{max}-\phi_0)\phi}{(\phi_{max}-\phi)\phi_0} \right]+ \log M,
\end{equation}
with $m$ and $M$ being parameters assumed to be linked to the hardening behavior and the compaction modulus, respectively.
This equation belongs to a new category of double logarithmic equation recently introduced \cite{Zhang2014}.

For binary mixtures, we use a simplified form of the equation proposed by Platzer et al. \cite{Platzer2018} as:
\begin{equation}
P = P_0(\kappa)\ln\left[\frac{\phi_{max}-\phi}{\phi_{max}-\phi^*(\kappa)}\right] + P^*,
\label{Eq:Platzer}
\end{equation}
with $P_0(\kappa)$ a characteristic pressure depending, {\it a priori}, on the proportion $\kappa$, $P^*$ a critical
pressure and $\phi^*(\kappa)$ a critical packing fraction at $P^*$.
Note that this equation was originally formulated in terms of void ratio
and developed in the context of a mixture of sand and rubber.
The rewriting of the Platzer et al. equation in terms of pressure versus packing fraction, simplified for a mixture of perfectly rigid particles with deformable ones is detailed in Appendix A.

Some compaction models are shown in Fig. \ref{fig:phi_p}, and their derivative following Eq. (\ref{Eq_Modulus}) are shown in Fig. \ref{fig:BULK}.
We see that they all fit well the compaction curves capturing as well the two horizontal asymptotes, i.e., one for the perfectly rigid granular assembly ($\phi\rightarrow \phi_0$), and the second for extremely high pressures ($\phi\rightarrow \phi_{max}$).
They also fit well the bulk evolution with the divergence observed as $\phi$ approaches to $\phi_{max}$,
although they slightly mismatch the evolution at higher pressures for all values of $\kappa$ and $\mu_s$.

In each case, these models require the measurement or calibration of many parameters.
By construction, for a given value of $\kappa$ and $\mu_s$, $\phi_0$  is known.
Then, following a non-linear least squares
regression, the maximum packing fraction $\phi_{max}$ and the
other fitting parameters involved in Eq. (\ref{Eq:Secondi}), Eq. (\ref{Eq:Zhang})
and Eq. (\ref{Eq:Platzer}) are simultaneously estimated in order to best fit the compaction curve.

\begin{figure}
    \centering
     \includegraphics[width=\linewidth]{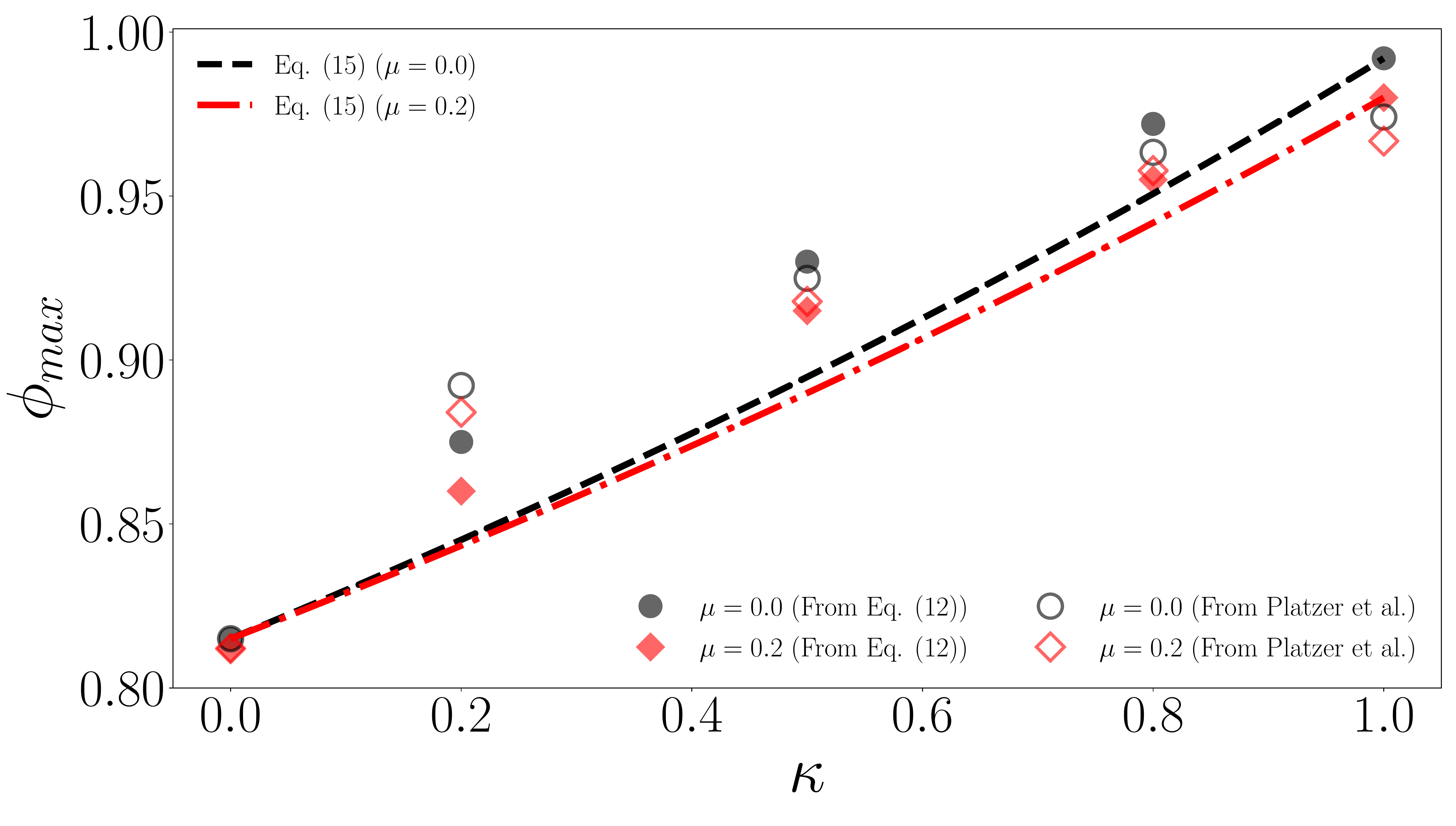}
    \caption{(Color online) Values of $\phi_{max}$ fitted with Eq. (\ref{Eq:Platzer}) (empty symbol) and with Eq. (\ref{eq:model}) (full symbol, see Sec. \ref{Sec_micromodels}) on the compaction curves of rigid-deformable particle assemblies shown in Fig. 3(a), for different $\kappa$ and $\mu_s\in\{0,0.2\}$.
    The dashed lines show the prediction given by Eq. (\ref{Eq_toy_model_max}).}
    \label{fig:phiMax}
\end{figure}

Now, for the assembly composed of only deformable particles
(i.e., for $\kappa=1$) we get
$n=m=1$, $(A_1,\phi_{max}) \simeq (M,\phi_{max}) \simeq (0.05E,0.97)$
for $\mu_s=0$, and $(A_1,\phi_{max}) \simeq (M,\phi_{max})\simeq (0.079E,0.96)$ for $\mu_s=0.8$.
For rigid-deformable particles assemblies, the best pair values of $P_0(\kappa)$ and $\phi_{max}$ in Eq. (\ref{Eq:Platzer}) are
summarized in Table \ref{tab:1} and Fig. \ref{fig:phiMax}, respectively,
imposing $P^*= 0$ and $\phi^*(k)=\phi_0$ for all values of $\kappa$ and $\mu_s$.
Note that Platzer et al. have shown that their model ceases to work for $\kappa>0.5$ while, in our case, the approximation is still acceptable even for $\kappa=1$.
Indeed, as discussed in Appendix \ref{A_Platzer}, the model of Platzer et al. requires knowing the evolution of the void ratio versus the pressure for a pure sand sample. This induces the fitting of more parameters.
\begin{table}
\centering
\begin{tabular}{c|c|c|c|c|c}
\hline\noalign{\smallskip}
    & $\kappa = 0.2$ & $\kappa = 0.5$ & $\kappa = 0.8$ & $\kappa = 1$  \\
\noalign{\smallskip}\hline\noalign{\smallskip}
$\mu_s=0.0$         & 0.17& 0.08 & 0.1  & 0.06  & \\
$\mu_s=0.2$      & 0.33& 0.11 & 0.1 & 0.075    &  \\
\noalign{\smallskip}\hline
\end{tabular}
\caption{Values of $P_0(\kappa)/E$ in Eq. (\ref{Eq:Platzer}) to best fit the compaction curves of rigid-deformable particle assemblies
shown in Fig. \ref{fig:phi_p}(a), as a function of $\mu_s$ and $\kappa$.}
\label{tab:1}
\end{table}

It is important to note different points. First, these models fit well our numerical data as long as well-chosen parameters are picked. Nonetheless, the physical meaning of these equations and of the induced parameters remain unclear.
Second, most of the existing equations are very similar. For example, the simplified form of the Platzer et al. equation is
equivalent to that of Secondi with $n=1$ and $A_1=P_0(\kappa)$. However, while $P_0(\kappa)$ is related to a characteristic pressure,
$A_1$ may be related to an equivalent yield stress according to \cite{Carroll1984}.
Third, the values we obtained for the fitting parameters in the equation of Secundi and Zhang are nearly equal ($A_1\simeq M$).
Finally, it is important to highlight that in all of the existing models,
the maximum packing fraction $\phi_{max}$ is estimated or calibrated along with the other parameters and cannot be easily deduced from existing equations.

A generalized compaction model for granular mixtures should, nonetheless, be based on a clear description of the mechanisms taking place at the scale of grains, their deformation, and their interactions.

\section{A micro-mechanical approaches}
\label{Sec_micromodels}

In granular assemblies, the compressive stress $P$ can also be deduced through the micro-mechanical expression of the granular stress tensor
defined by \cite{Andreotti2013}:
\begin{equation}
\sigma_{ij} =  \frac{1}{V}  \sum_{c \in V} f_i^c \ell_j^c = n_c \langle f^c_i\ell^c_j \rangle_c,
\label{eq:sigma}
\end{equation}
where $f_i^c$ is the $i$ component of the contact force acting on a contact $c$ and $\ell_j^c$ is the
$j$ component of the branch vector (i.e., the vector joining the centers of particles interacting at contacts $c$).
The sum runs over all contacts inside the volume $V$, and $\langle...\rangle_c$ is the average over contacts.
The density of contact $n_c$, on the right hand side of Eq. (\ref{eq:sigma}), is given by $n_c=N_c/V$ with $N_c$ the total number of contacts in the volume $V$.
From the stress tensor, we extract the mean stress $P_\sigma = (\sigma_1+\sigma_2)/2$, with $\sigma_1$ and $\sigma_2$ the principal stress values, and $P=P_\sigma$.

Considering a small particle size dispersion around the diameter $\langle d \rangle$, the contact density can
be rewritten as $n_c = 2\phi Z / \pi \langle d \rangle^2$, with $Z=2N_c/N_p$ the coordination number.
These definitions permit to rewrite the stress tensor as $\sigma_{ij} = (2\phi Z/\pi \langle d \rangle^2) \langle f_i^c \ell_j^c \rangle_c$.
Finally, taking into account the definition of $P$ via the principal stresses of $\sigma_{ij}$, we can deduce a
microstructural equation of the compressive stress as \cite{Agnolin2007c,Khalili2017b,Bathurst1988}:
\begin{equation}\label{eq:Pglobal_local}
P = \frac{\phi Z}{\pi} \sigma_{\ell},
\end{equation}
with $\sigma_{\ell} = \langle f^c \cdot \ell^c \rangle_c/\langle d \rangle^2$, a measure of the inter-particle stresses, with $\cdot$ the scalar product.

Equation \ref{eq:Pglobal_local} reveals the mutual relation between $P$ and $\phi$ through the granular microstructure
described in terms of both particle connectivity ($Z$) and inter-particle stress ($\sigma_\ell$).

\subsection{Particle connectivity}
The coordination number $Z$, allowing to quantify the average number of neighbors per particle,
is the first and the simplest statistical descriptor of the granular texture, i.e., the organization of the particles and their contacts in space.

At the jammed state (i.e., for the packing fraction $\phi_0$), the packing is characterized by a minimal value $Z_0$.
Below such value, the collective movement of the particles is possible without implying particle deformation.
As described in several earlier studies, $Z_0$  depends on many parameters like shape, friction and packing preparation \cite{Hecke2009_Jamming,Silber2002_Geometry,Donev2005_Pair,Khalili2017a} to name a few.
Basically, for circular particle assemblies, $Z_0$ declines with $\mu_s$ and tends to $4$ as $\mu_s\rightarrow 0$, and to $3$ for large friction values \cite{Agnolin2007a}.
Furthermore, since $Z_0$ also depends on the packing preparation for frictional particles, distinct values of $Z_0$ are admissible.

Above the jammed state, it has been systematically reported in the literature that $Z$ continues to increase following a power-law as:
\begin{equation}
\label{Eq_Z_Phi}
Z-Z_0 = \xi (\phi-\phi_0)^\alpha,
\end{equation}
with $\alpha\sim0.5$ and $\xi=(Z_{max}-Z_0)/(\phi_{max}-\phi_0)^\alpha$ a structural parameter fully defined as $P/E\rightarrow \infty$, with both $\phi$ and $Z$ reaching a maximum value $\phi_{max}$ and $Z_{max}$, respectively.
This relation was observed both numerically and experimentally for many deformable particulate assemblies like foams, emulsions, and rubber-like particles \cite{Vu2019,Durian1995_Foam,Katgert2010_Jamming,Majmudar2007_Jamming}.

\begin{figure}
    \centering
   \includegraphics[width=\linewidth]{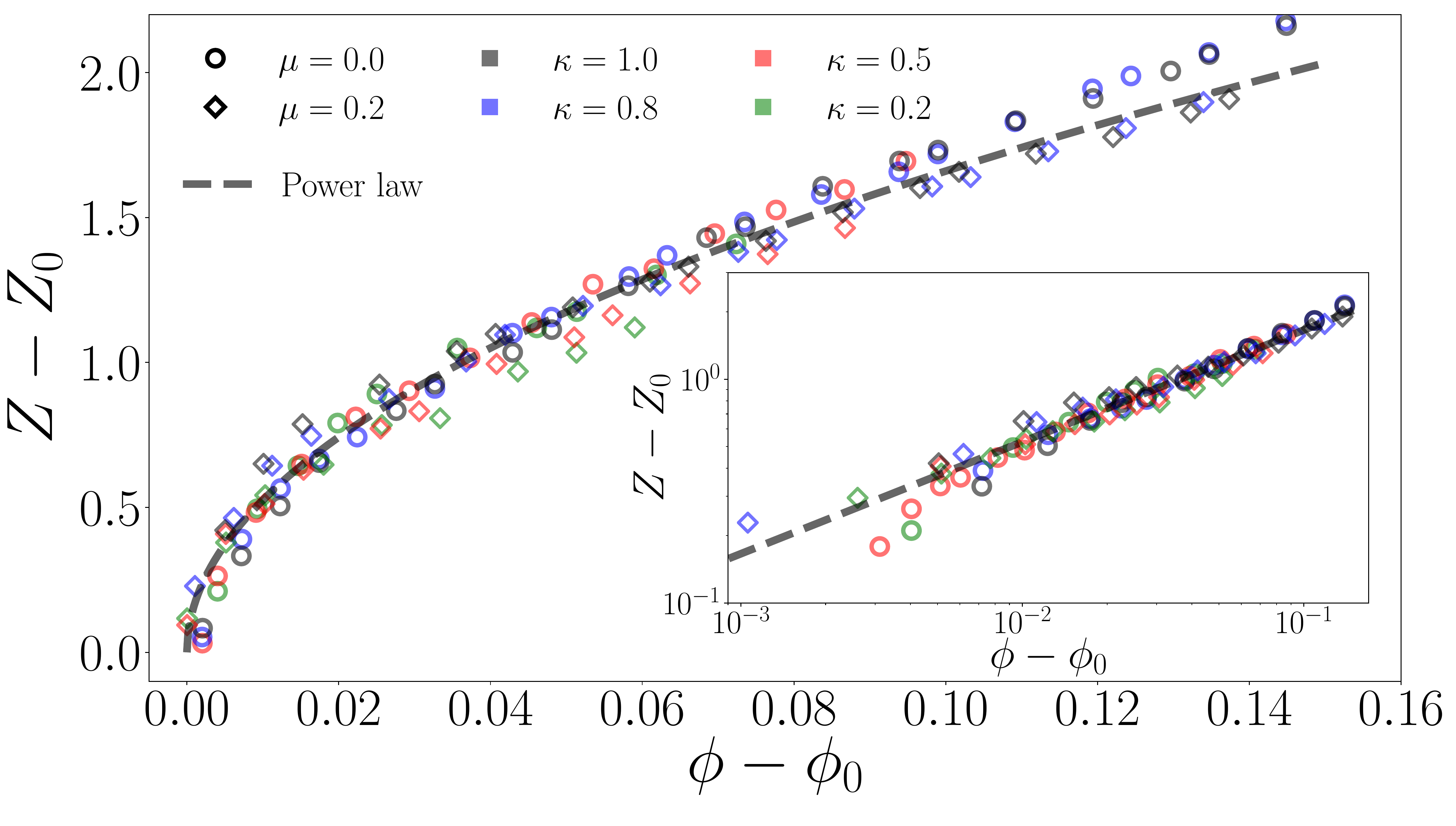}
    \caption{(Color online)  The reduced coordination number $Z-Z_0$ as a function of the reduced solid fraction $\phi - \phi_0$
    for all values of $\kappa$ and $\mu_s$ (a log-log representation is shown in the inset).
    The dashed black line is the power-law relation \quad \quad  $Z-Z_0 = \xi (\phi-\phi_0)^\alpha$ with $\alpha=0.5$ and $\xi = 5.1$.}
    \label{fig:z_phi_Z}
\end{figure}

As shown in Fig. \ref{fig:z_phi_Z}, we found the same proportionality in our simulations, with $\xi\sim 5.1$,
independently of the mixture ratio and friction.
Thus, our results extend the validity of such relation to binary mixtures.

\subsection{Elastic Modulus}
Moreover, the inter-particle stress $\sigma_\ell$ could be related to the packing fraction,
either by considering deformations at the contact, or through the
bulk properties of an elementary system composed of a single elastic particle.

\subsubsection{Voigt approximation}
Elastic properties of a granular assembly can be estimated in the small-strain domain through the
Voigt approximation (also called effective medium theory EMT \cite{Bathurst1988,Khalili2017a,Makse1999,Magnanimo2012}), in which the particles
are replaced by a network of bonds of length $\ell^c$. From there, and by analogy with the macroscopic volumetric
strain $\varepsilon_v$, we can  define a local volumetric strain  as $\varepsilon_{v,\ell}=2 \ln (\langle \ell^c\rangle /\langle d\rangle)$.
Our numerical simulations also show that  $\varepsilon_{v}=2\varepsilon_{v,\ell}$, for all values of $\kappa$ and $\mu_s$.
Then, we assume that the inter-particle stress between two deformable particles, or between a deformable and rigid
particle, is given by  $\sigma_\ell = E \varepsilon_{v,\ell}$.

Finally, the above expressions with Eqs. \ref{Eq_Z_Phi}, \ref{eq:Pglobal_local}, and \ref{Eq_Modulus} together with a first-order Taylor expansion of
$\varepsilon_v$, give an estimation of the bulk modulus as:
\begin{equation}
\label{Eq_Bulk_K1}
 \frac{K_1}{E}= \frac{ Z\phi}{2\pi} \left(\frac{5}{2} - \frac{\phi_0}{\phi}\right) - \frac{Z_0\phi}{4\pi}.
\end{equation}
Note that, in the limit of $\phi\rightarrow\phi_0$, Eq. (\ref{Eq_Bulk_K1}) predicts that
$K_1\rightarrow Z\phi E/(2\pi)$
which is in agreement with other equations obtained within a small-strain framework  for assemblies of rigid particles with elastic
interactions \cite{Khalili2017a,Makse1999,Zaccone2011_Approximate,Magnanimo2012}.

\begin{figure}
    \centering
    \includegraphics[width=\linewidth]{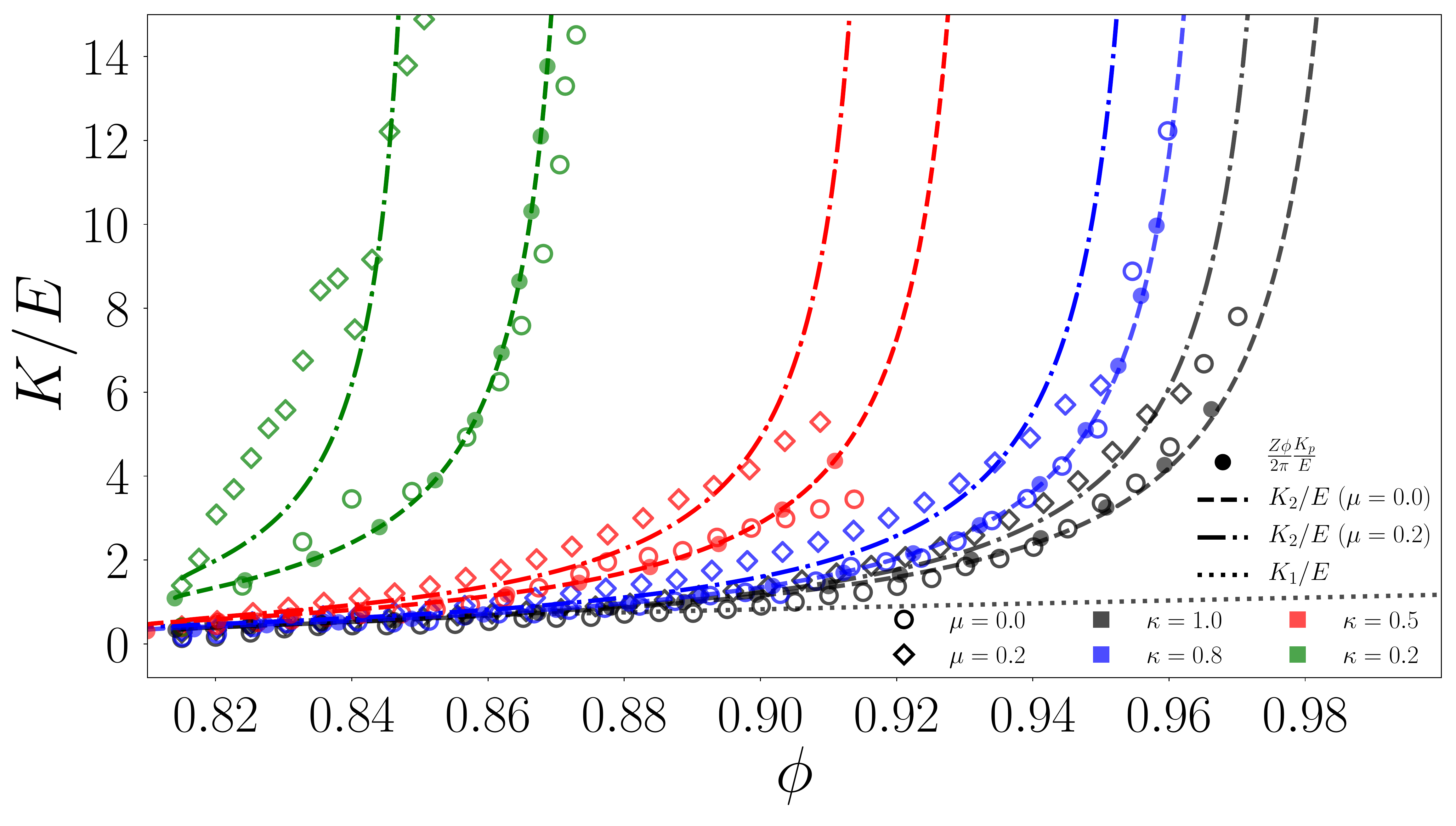}
    \caption{(Color online)  Bulk modulus $K$ normalized by $E$ (empty symbols) along with
    the micro-mechanical relation proposed on Eq. (\ref{eq:K_Kp}) (full symbols)
    for all values of $\kappa$ and $\mu_s=0$.
    The prediction given by  Eq. (\ref{eq:K_predict}) is shown in dashed line for $\mu_s=0$ and $\mu_s=0.2$
    and the one given by Eq. (\ref{Eq_Bulk_K1}) is displayed in dotted line.}
    \label{fig:K_prediction}
\end{figure}

The prediction given by Eq. (\ref{Eq_Bulk_K1}) is shown in Fig. \ref{fig:K_prediction} for all values of $\kappa$ at $\mu_s=0$.
We see that the measurement of the bulk modulus within an equivalent medium approach gives matching results in the small-strain domain for all values of $\kappa$. However, here the prediction given by Eq. (\ref{Eq_Bulk_K1}) is still acceptable over the range of $\phi\in[\phi_0,\phi^+]$ where
$\phi^+$ increases from $\simeq0.83$ for $\kappa=0.2$ to $\simeq0.9$  for $\kappa=1.0$,  until an
increasing mismatch is observed as the packing fraction tends to $\phi_{max}$.  Indeed, in the limit $\phi\rightarrow\phi_{max}$
the assembly of grains starts to behave as a non-deformable solid, and thus, the corresponding bulk modulus diverges.
These observations suggest that the definition of local strains only by means of the contact deformations $\sigma_\ell = E \varepsilon_{v,\ell}$ should be reconsidered.

\subsubsection{Scaling with a single particle configuration}
\begin{figure}
    \centering
    \quad \quad \quad \includegraphics[width=0.8\linewidth]{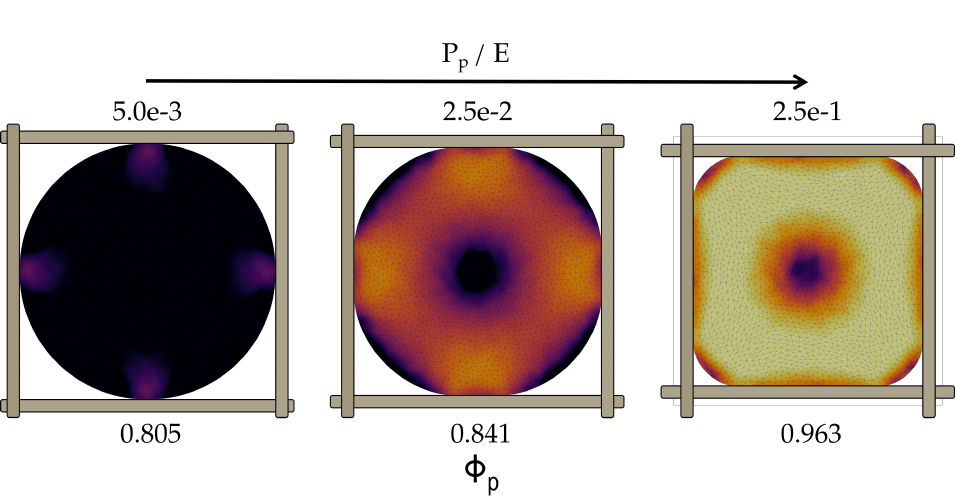}
    \includegraphics[width=0.9\linewidth]{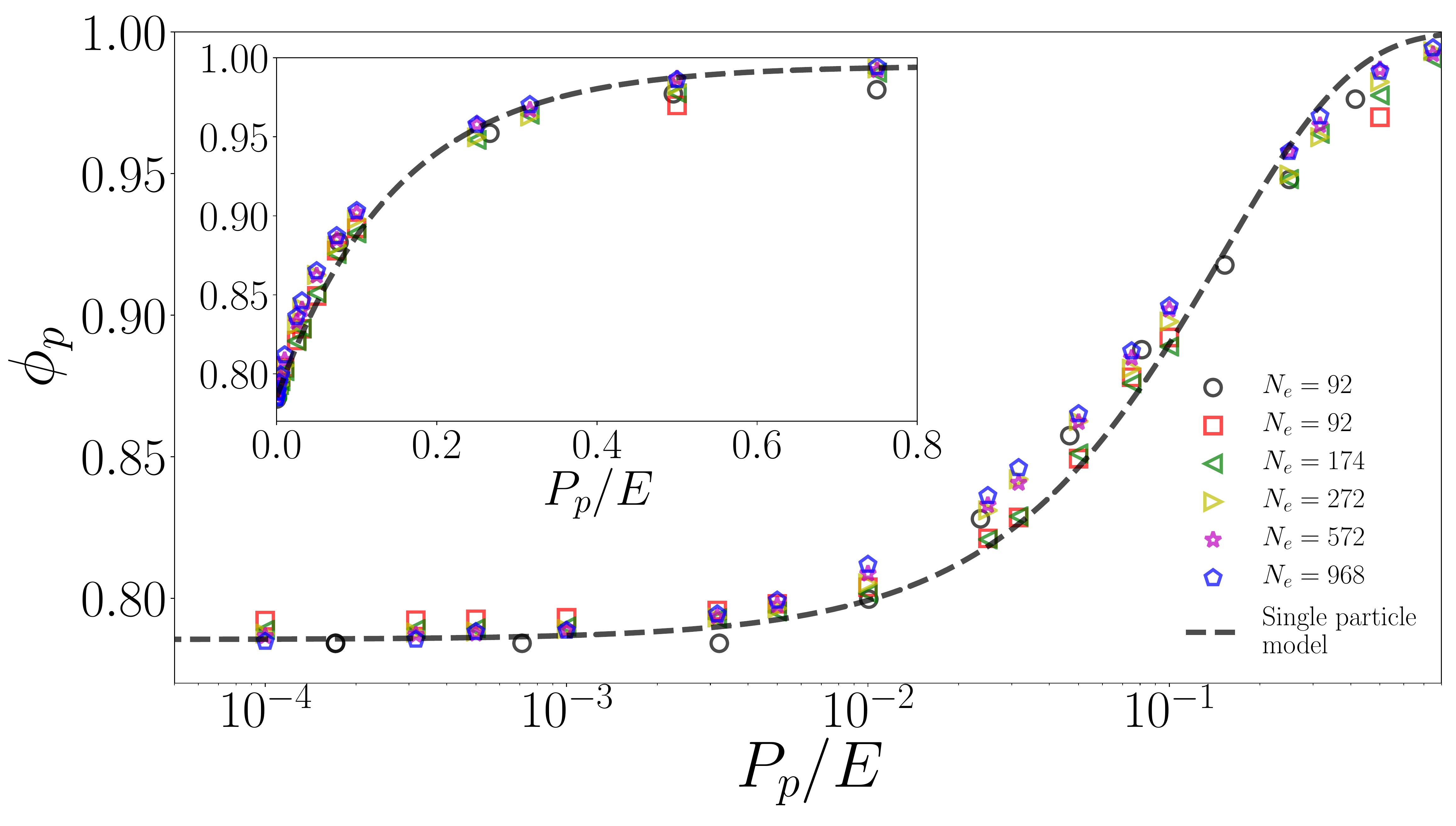}
    \caption{(Color online)  Compaction curve of a single particle inside a square box for different mesh resolutions. a: Snapshots for $N_e = 968$ of the simulation at different compression levels. The color intensity of the particle is proportional to the volumetric deformation. b: Packing fraction as a function of the scaled pressure applied on the particle.
    Red squares ($P_p$ fixed and $E$ varied) and black circles ($E$ fixed and $P_p$ varied) are tests on a particle with
    $N_e=92$ finite elements. For the other mesh resolutions, $E$ was fixed and $P_p$ varied.}
    \label{fig:Pp_vs_Phip}
\end{figure}

In this section, a different point of view is proposed.
Let us consider the case of an elementary system composed of a single particle isotropically compressed between four rigid walls
(i.e., submitted to the same boundary conditions as the multi-particle assembly, see the upper part of Fig. \ref{fig:Pp_vs_Phip} ).
In Fig. \ref{fig:Pp_vs_Phip}, we present the evolution of the packing fraction $\phi_p$ as a function of the applied pressure $P_p$ for the single particle case.
Note that the number of finite elements $N_e$ has a small influence on the results.
We also observe that the single particle compression curve $\phi_p-P_p$ is roughly similar to the
multi-particle compaction curve $\phi-P$ (Fig. \ref{fig:phi_p}).
This supports the idea of a strong relation
between the single particle and multi-particle systems.

Indeed, as shown in Fig. \ref{fig:Pp_vs_Phip} (gray dashed line),
the compaction behavior of such elementary system is well described with the following logarithmic function:
\begin{equation}
\label{eq:evol_1P}
P_p/E = - b \ln \left(\frac{\phi_{p,\mathrm{max}}-\phi_p}{\phi_{p,\mathrm{max}}-\phi_{p,0}}\right),
\end{equation}
with $\phi_{p,\mathrm{max}}$ the maximum packing fraction obtained, $\phi_{p,0}=\pi/4$ the solid fraction as $P_p/E\rightarrow 0$,
and $b$ a constant of proportionality found to be $\simeq0.14$.
Equation \ref{eq:evol_1P} is derived from the analogy to the collapse of a cavity within an elastic medium under isotropic compression
following Carroll et al. \cite{Carroll1984}.
Although this relation is well adapted to the single particle test, similar functional forms have been used for multi-particle systems, as discussed in Sec. \ref{Discussion_models}.
Then, the bulk modulus of the single particle assembly is given by:
$K_p(\phi_p) = (dP_p/d\phi_p) \cdot (d\phi_p/d\varepsilon_{v,p})$, with $\varepsilon_{v,p} = -\ln(\phi_{p,0}/\phi_{p})$.

Now, comparing these two systems at equivalent packing fraction (i.e., for $\phi_p\equiv\phi$),
we obtain that, for all values of $\kappa$ and $\mu_s$, the
macroscopic bulk modulus $K$ of the assembly scales with $K_p$ as (see full symbols in Fig. \ref{fig:K_prediction}):
\begin{equation}\label{eq:K_Kp}
K \equiv \frac{Z\phi}{2\pi} K_p + \mathcal{O},
\end{equation}
with $\mathcal{O}$ negligible high order terms on $\phi$.
Equation \ref{eq:K_Kp} allows us to reinterpret the micromechanical origin of the bulk modulus of an assembly of
rigid-deformable particles in terms of particle connectivity,
packing fraction, and the bulk property of an elementary system.
We can also reinterpret Eq. (\ref{eq:Pglobal_local}) as:
\begin{equation}\label{eq:P_Pp}
P \simeq \frac{Z\phi}{2\pi} P_p,
\end{equation}
and we deduce that $\mathcal{O}$ in Eq. (\ref{eq:K_Kp}) is related to the derivatives $(dZ\phi P_p+Zd\phi P_p)$.
Equations \ref{eq:K_Kp} and \ref{eq:P_Pp} reveal that the
elastic and compaction properties of a binary mixture are scalable from the behavior of
a single particle.
This finding is aligned with the general idea of describing the macroscopical properties of a granular packing from a single representative element \cite{Tighe2011_Stress,Cardenas2018_Contact}.

\subsection{A compaction and bulk equation}
Finally, introducing the $Z-\phi$ relation (i.e., Eq. (\ref{Eq_Z_Phi})) into Eq. (\ref{eq:P_Pp}), together with Eq. (\ref{eq:evol_1P})
at equivalent packing fraction, and noting that, for a given friction the maximum packing fraction depends on mixture ratio (Fig. \ref{fig:phiMax}),
we get the following compaction equation:
\begin{equation}\label{eq:model}
\frac{P(\phi,\kappa)}{E} \simeq -\frac{b\phi}{2\pi}\left\{Z_0 + \xi(\phi-\phi_0)^{\alpha}\right\} \ln \left(\frac{\phi_{\mathrm{max}}(\kappa)-\phi}{\phi_{\mathrm{max}}(\kappa)-\phi_0}\right).
\end{equation}
In contrast to previous models, the only unknown parameter in  Eq. (\ref{eq:model}) is
the maximum packing fraction $\phi_{max}(\kappa)$ since
all other constants are determined from either the initial state, the behavior of a single representative particle, or the mapping between the packing fraction and coordination curve.

\begin{figure}
    \centering
    \includegraphics[width=\linewidth]{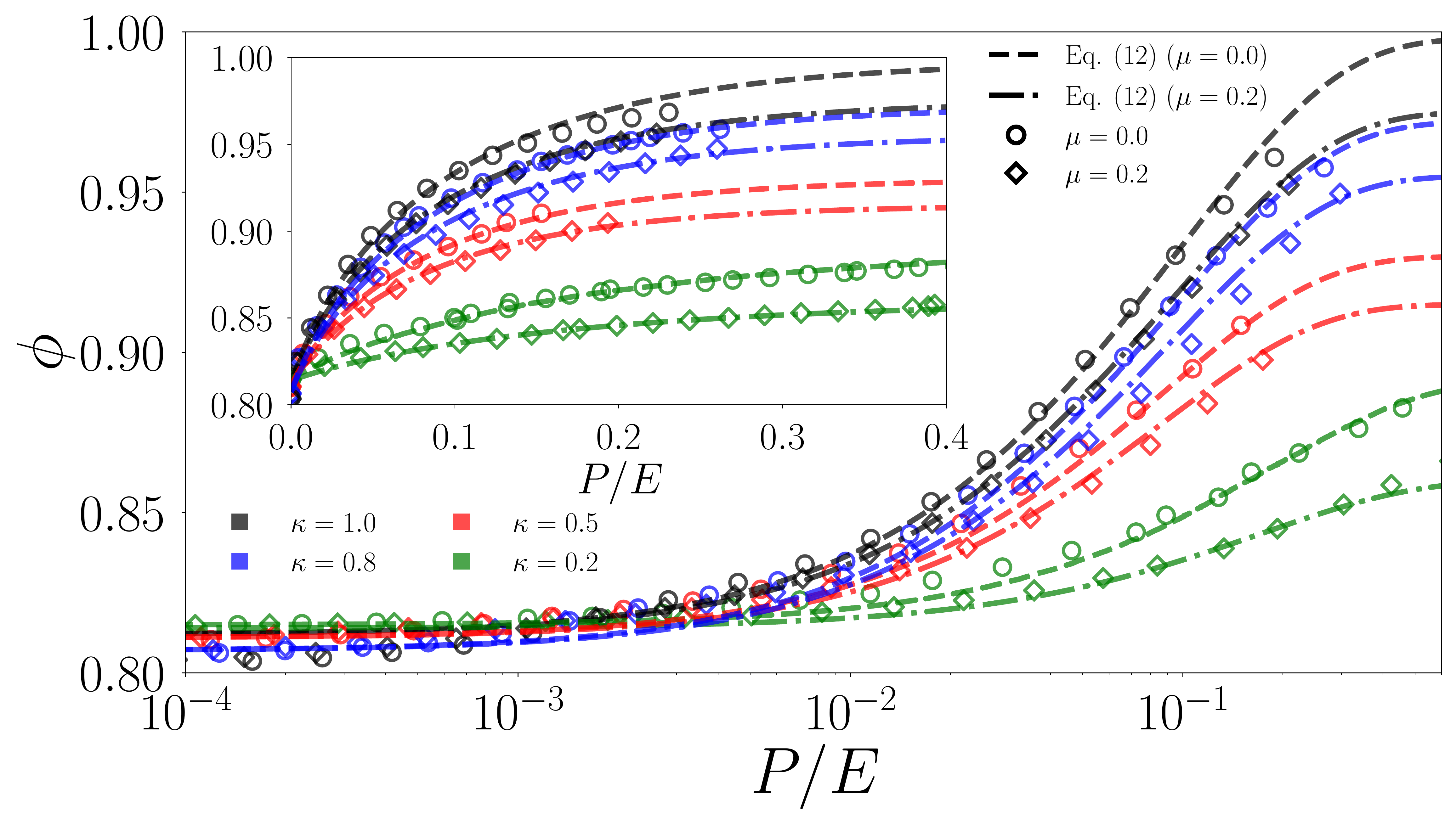}(a)
    \includegraphics[width=\linewidth]{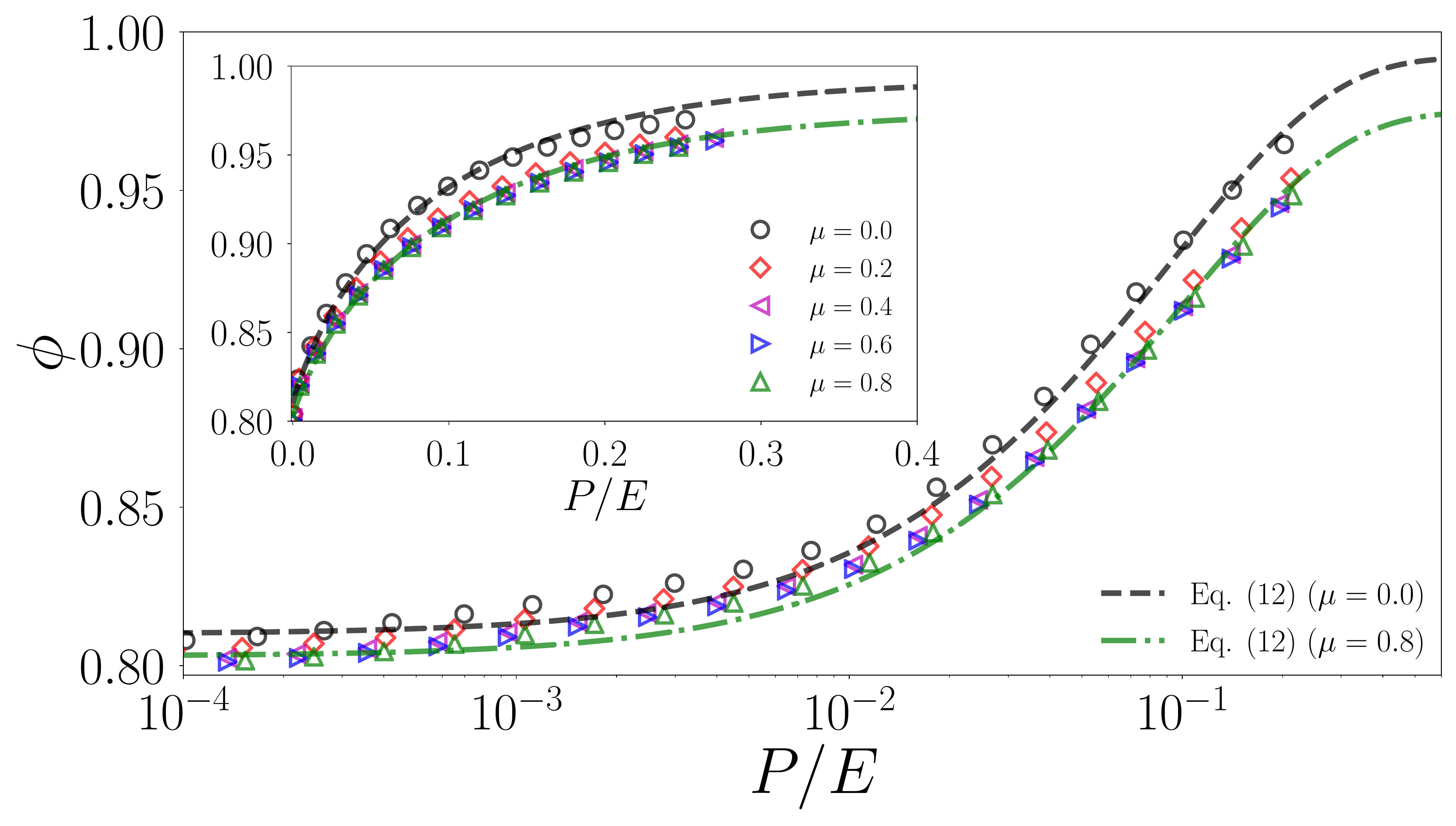}(b)
    \caption{(Color online)  Compaction curves $\phi$ as a function of $P/E$ for
    (a) rigid-deformable particles assemblies with $\kappa \in[0.2,...,1]$ and $\mu_s\in\{0,0.2\}$, and for (b) completely deformable particle assemblies  (i.e., $\kappa=1$) with $\mu_s\in\{0, ... 0.8\}$. In both, the curves are presented in lin-log scale in the main panel and in lin-lin scale in the inset. The predictions given by our micro-mechanical equation (Eq. (\ref{eq:model})) are shown in dashed lines.}
    \label{fig:P_phi_prediction}
\end{figure}

Figure \ref{fig:P_phi_prediction} presents our numerical data (same as in Fig. \ref{fig:phi_p}) together with the compaction equation
given by Eq. (\ref{eq:model}).
We see that the predictions given by Eq. (\ref{eq:model}) are
outstanding for any pressure capturing the asymptotes
for vanishing and extremely high pressures, the effect of mixture ratio, and the effect of friction.
Our compaction equation also allows us to predict the asymptote for the maximal packing fraction $\phi_{max}$,
both as a function of $\kappa$ and $\mu_s$.
The best $\phi_{max}$-values used in Eq. (\ref{eq:model}) are shown in Fig. \ref{fig:phiMax} (full symbols) as a function of $\kappa$ for
$\mu_s\in\{0,0.2\}$.

Going one step further and derivating Eq. (\ref{eq:model}) we get a second expression for the bulk modulus evolution as:
\begin{equation}
\label{eq:K_predict}
\frac{K_2(\phi,\kappa)}{E} = \frac{b\phi^2}{2\pi(\phi_{\mathrm{max}}(\kappa)-\phi)} \{Z_0 + \xi(\phi-\phi_0)^{\alpha}\}.
\end{equation}
Figure \ref{fig:K_prediction} shows the above relation giving the evolution of $K$ throughout the
deformation for all values of $\kappa$ and $\mu_s\in\{0,0.2\}$.

\begin{figure}
    \centering
    \includegraphics[width=0.7 \linewidth]{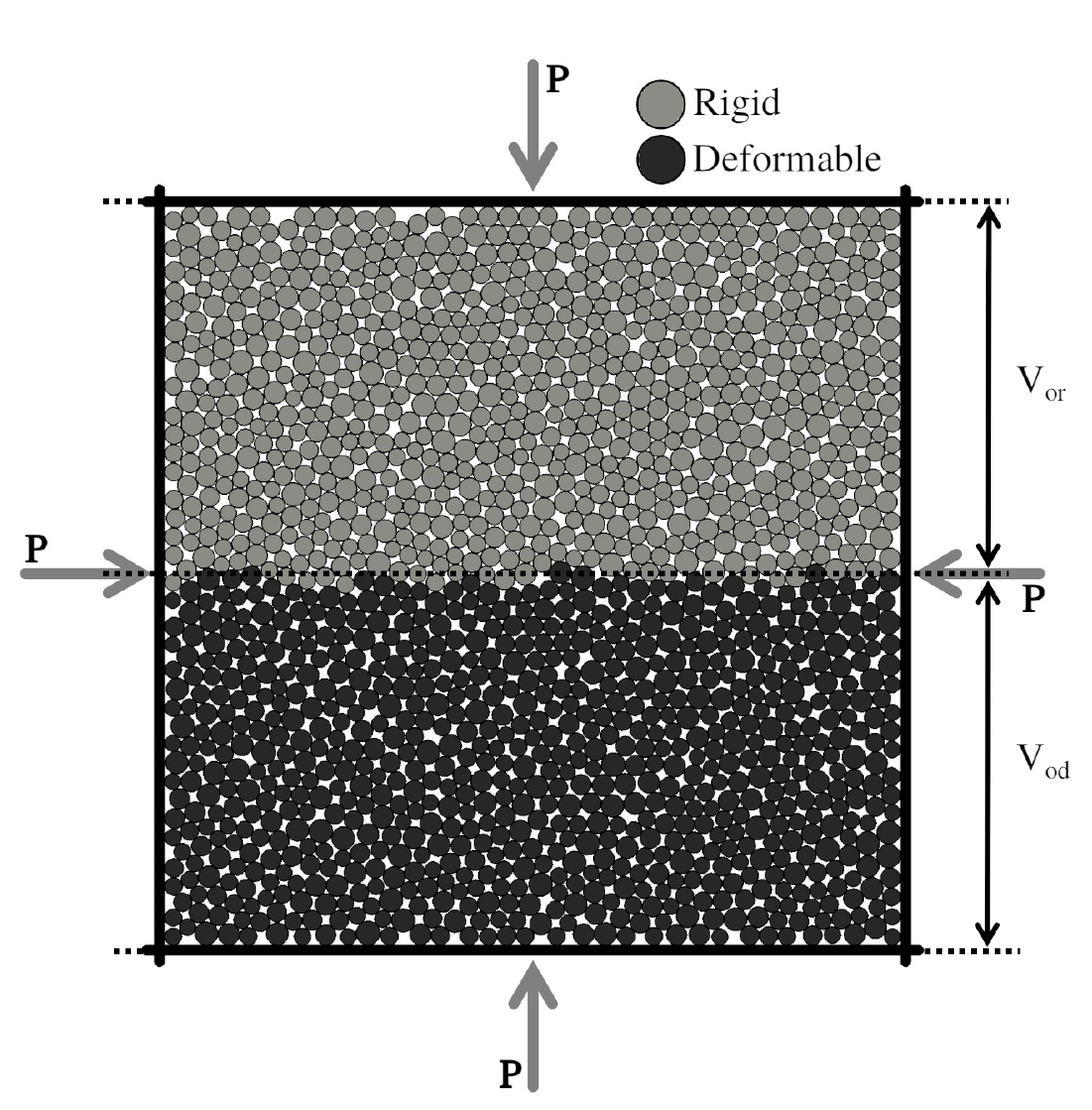}
    \caption{(Color online) Sketch of the simplified geometrical approach to estimate the packing fraction of the assembly
    as a function of $\kappa$. Particles are separated in two distinct phases, a rigid and a deformable one.}
    \label{fig:toy_model}
\end{figure}

Finally, the increase of $\phi_{max}$ with $\kappa$, shown in Fig. \ref{fig:phiMax}, can be captured considering a simple system.
Let us imagine an assembly where particles are separated in two phases, a rigid and a deformable one,
as shown in Fig. \ref{fig:toy_model}.
The total volume $V$ of the box is then given by $V=V_{or}+V_{od}$, where $V_{or}$ and $V_{od}$
are the corresponding volumes of the sub-boxes containing the rigid and deformable particles respectively.
Considering that the total volume of deformable particles is $V_d=\kappa V_p$ and that the one of rigid particles
is $V_r=(1-\kappa) V_p$,  the packing fraction of such demixed mixture is then given by:
\begin{equation}
\label{Eq_toy_model}
\phi(\kappa) = \frac{\phi_0\;\phi(1) }{\phi(1)+(\phi_0-\phi(1))\kappa},
\end{equation}
with $\phi_0=V_r/V_{or} = \phi(\kappa=0)$, and
$\phi(1) =V_d/V_{od}= \phi(\kappa=1)$, being the packing fractions of only rigid and deformable particle assemblies, respectively.
Therefore, we can write the maximal packing fraction of the mixture as
\begin{equation}
\label{Eq_toy_model_max}
\phi_{max}(\kappa) = \frac{\phi_0\;\phi_{1,max} }{\phi_{1,max}+(\phi_0-\phi_{1,max})\kappa},
\end{equation}
with $\phi_{1,max}$ the maximum packing fraction at $\kappa=1$.
As shown in Fig. \ref{fig:phiMax} with dashed lines, Eq. (\ref{Eq_toy_model}) gives acceptable
predictions for the evolution of $\phi_{max}$ as $\kappa$ increases both
for $\mu_s=0$ and $\mu_s=0.2$.
So, by replacing Eq. (\ref{Eq_toy_model_max}) into Eq. (\ref{eq:model}) and Eq. (\ref{eq:K_predict}),
we obtain predictive equations based on the sole knowledge of the maximum compaction value
attainable in assemblies composed of only deformable particles.

\section{Conclusions and perspectives}
\label{Sec_conclu}
In summary, by means of extensive non-smooth contact dynamics simulations, we analyze the compression behavior of bidimensional granular assemblies composed of mixtures of rigid and incompressible deformable particles. The deformable bodies are simulated following a hyper-elastic
neo-Hookean constitutive law using classical finite elements. The proportion of deformable particles was varied from $0.2$ (i.e., assembly composed of 20\% of deformable grains) to $1$ (i.e., assembly composed of only deformable grains) for different values of the friction.
Starting from the jammed state characterized by a packing fraction $\phi_0$, packings were isotropically compressed by gradually applying stress on the boundaries.

We found that, for all values of the mixture ratio and friction, the packing fraction increases from $\phi_0$ and
asymptotically tends to a maximum value $\phi_{max}$.
We showed that $\phi_{max}$ decreases as the proportion of deformable particles declines, and the friction coefficient increases.
Although most of the existing models provide acceptable predictions, the maximum packing fraction reachable must be estimated or calibrated along with other parameters and cannot be easily deduced.

A major outcome of this work is the introduction of a compaction model for binary mixtures of
rigid and deformable particles, free of {\it ad hoc} parameters, and standing on well-defined
 quantities related to 1) particle connectivity, 2) the bulk behavior of a single
representative particle and 3) the proportion of rigid-deformable particles.
Our model, derived from the micro-mechanical expression of the granular stress tensor, results in outstanding predictions of the compaction evolution for all values of the mixture ratio parameter at any friction.
In addition to this, the maximum packing fraction of the assembly is deduced as the only fitting parameter at the macroscopic scale.
From the compaction model, a bulk equation is also deduced, resulting in good
agreement with our numerical data for all values of the mixture ratio parameter and  friction.

On top of the obtained compaction equation for the assembly of rigid-deformable particles, this paper highlights the methodology via a micro-mechanically approach.
This approach allows us to unify in a coherent framework the compaction behavior of
assemblies of deformable and rigid-deformable grains beyond the jamming point.
The above framework may now be used and extended to analyze much more complex
deformable granular assemblies
by considering a wide range of material properties, such as plastic particles, non-spherical particles, and polydisperse particles.
Other loading configurations, like the oedometric compression test, could be investigated in this framework as well.
These alternative mechanical considerations may lead to distinct functional forms,
but certainty, settled on the behavior of a single representative particle and the
evolution of the packing connectivity.

Finally, we recall that the granular stress tensor, from which the model is built,
is an arithmetic mean involving the branches and contact force vectors.
Thus, high order statistical descriptions of these quantities (other than just their average value)
should be considered and will allow us to characterize in a finer way the granular texture, its evolution as well as the force and the stress transmission beyond the jamming point. \\

We thank Frederic Dubois for the valuable technical advice on the simulations in LMGC90. We also acknowledge the support of the High-Performance Computing Platform MESO@LR.

\appendix

\section{rewriting of the Platzer et al. model}
\label{A_Platzer}
Through the literature review, we note that the majority of the compaction models were introduced in the form of $P(\phi)$.
In contrast, Platzer et al. \cite{Platzer2018}, by studying a mixture of sand and rubber, introduced their model in the form of  $e(\kappa,P)$,
with $e$ the void fraction of the assembly defined by $e=1-\phi$.
Assuming that the void space is fulfilled following a first-order function of the applied pressure, they proposed the following compaction equation:
\begin{equation} \label{eq:PlatzerOriginal}
    e(\kappa,P) = e(0,P) - \kappa f^* -
      \kappa (F-f^*)\left[1-\exp\left(\frac{P^*-P}{P_0(\kappa)}\right)\right] ,
\end{equation}
with $e(0,P)$ the experimental void fraction at pressure $P$ for sands,
$P_0(\kappa)$ and $P^*$ two characteristic pressures.
$f^*=f(\kappa,P^*)$ and $F=f(\kappa,P\rightarrow \infty)$ are the critical deformed fraction and
the maximum deformed fraction of rubber, respectively. They are defined from:
\begin{equation} \label{eq:Platzerf}
    f(\kappa,P) = \frac{e(0,P) - e(\kappa,P)}{\kappa}
\end{equation}

Replacing Eq. (\ref{eq:Platzerf}) into Eq. (\ref{eq:PlatzerOriginal}), and after some algebra,
the equation of Platzer et al. can be rewritten as:
\begin{equation} \label{eq:PlatzerModified1}
    P(\kappa,\phi) = -P_0(\kappa)
     \ln \left(\frac{[\phi_{max}-\phi]+[\phi(0)-\phi_{max}(0)]}
    {[\phi_{max}-\phi^*(\kappa)]+[\phi^*_{0}(0)-\phi_{max}(0)]}\right) + P^*,
\end{equation}
with $\phi_{max}=\phi_{max}(\kappa)$, $\phi(0)=\phi(\kappa=0,P)$
the packing fraction evolution for a pure sand sample, $\phi_{max}(0)=\phi_{max}(\kappa=0)$
the maximum packing fraction obtained for a pure sand sample,
and $\phi^*_{0}(0)=\phi^*_{0}(\kappa=0)$ and $\phi^*(\kappa)$ the packing fraction
for a pure sand and a mixture rubber-sand, respectively, at a given initial confining pressure $P^*$.

Now, considering that the grains of sand are replaced by perfectly rigid particles, and starting the compaction process at the jammed state, we naturally get that, for $\kappa=0$, $\phi(0)=\phi^*_{0}(0)=\phi_{max}(0)$.
This leads to a simplified version of the equation of Platzer et al. as:
\begin{equation} \label{eq:PlatzerModified2}
    P(\kappa,\phi) = -P_0(\kappa) \ln\left(\frac{\phi_{max}(\kappa)-\phi}
                                {\phi_{max}-\phi^*(\kappa)}\right) + P^*.
\end{equation}

\bibliographystyle{apsrev}
\bibliography{biblio}

\end{document}